\documentclass[
    aps,
    fleqn,
    a4paper,
    superscriptaddress,
    preprint,
    preprintnumbers,
    showpacs,
    floatfix,
    showkeys
]{revtex4}

\usepackage{graphicx}
\usepackage{tabularx}
\usepackage{supertabular}
\usepackage{amsmath}
\usepackage{amsfonts}
\usepackage[cp1250]{inputenc}
\usepackage{color}
\usepackage[
    colorlinks,
    linkcolor=blue,
    anchorcolor=blue,
    citecolor=blue,
    urlcolor=blue,
    menucolor=blue,
    filecolor=blue
]{hyperref}

\newcommand{\1}[1]{#1^{(1)}}
\newcommand{\2}[1]{#1^{(2)}}

\begin{document}

\title{A method to study ageing of polydomain ferroelectrics using measurements of nonlinear permittivity}

\author{Pavel \surname{Mokr\'{y}}}
\email{pavel.mokry@tul.cz} 
\affiliation{Institute of Mechatronics and Computer Engineering, Technical University of Liberec, CZ-46117 Liberec, Czech Republic}

\date{\today}

\begin{abstract}
It is known that the permittivity of the ferroelectric films is affected by 
several phenomena, which deteriorate the material quality (e.g. the 
redistribution of the crystal lattice defects, appearance of the 
electrode-adjacent non-ferroelectric layers or the spontaneous polarization 
screening due to a free charge injection across the electrode-adjacent layer, 
etc.). It is also known that the permittivity of ferroelectric polydomain 
films is controlled by the sum of two contributions: the crystal lattice 
(intrinsic) contribution and the domain wall movement (extrinsic) 
contribution. It is the latter one, which is very sensitive to the 
aforementioned phenomena and which plays a key role in the deterioration of 
the dielectric response of the ferroelectric polydomain films. In this Article,
 there is presented a method for the identification of the process, which is 
responsible for the ferroelectric ageing. The method is based on the analysis 
of the evolution of both the linear and nonlinear permittivity during ageing. 
Applicability of the method is theoretically demonstrated on four ageing 
scenarios in two qualitatively different systems where the evolution of the 
nonlinear permittivity is controlled, first, by a redistribution of the 
pinning centers on the domain wall and, second, by microstructural changes at 
the interface between the ferroelectric layer and the electrode. It is shown 
that each ageing scenario is characterized by unique trend in the evolution 
between the linear and nonlinear part of the permittivity, which can be 
verified experimentally.
\end{abstract}



\maketitle

\section{Introduction}
\label{sec:Intro}

Ferroelectric materials play more and more important role in many contemporary 
electronic devices. Due to the fundamental thermodynamic reasons, researchers 
and applied scientists integrating the ferroelectric materials into 
silicon-based semiconductor chips often face problems with a severe 
deterioration of the quality of the integrated ferroelectric films. These 
deterioration problems may manifest themselves during characterization 
experiments such as ferroelectric hysteresis loop measurements, pulse 
switching curve measurements, or small signal permittivity measurements, and 
cause phenomena called imprint, fatigue, ageing, etc. Unfortunately, 
determination of the ``microstructural'' mechanism predominantly responsible 
for an observed deteriorating effect in a particular ferroelectric sample is 
always very difficult, since several different microstructural mechanisms 
yield the similar manifestations in the aforementioned experiments. In order 
to get a better insight into microstructural mechanisms in ferroelectric 
samples, which would serve as a feedback for the optimization of the fabrication 
processes, a comprehensive analysis of the results of several characterization 
techniques should be performed.

The aforementioned issues have motivated the study presented in this work, 
where we analyze the possibility to use the measurements of nonlinear 
permittivity for the determination of mechanisms responsible for the 
ageing of polydomain ferroelectrics. The principle of our study is based on the 
fact that the dielectric response of the polydomain ferroelectric film is 
controlled by the sum of two contributions: the crystal lattice (intrinsic) 
contribution and the domain wall movement (extrinsic) contribution. Since the 
latter one is known to be sensitive to the geometry of the domain pattern and to 
several microstructural features of the material such as crystal 
lattice defects or free charges, which may work as pinning centers and may 
reduce the mobility of the domain walls, it is very likely that a 
comprehensive analysis of nonlinear dielectric permittivity may offer a simple 
and useful tool for getting some additional information about the 
microstructural processes responsible for the ferroelectric ageing.

The concept of the ferroelectric domain boundary contribution to permittivity 
has been discussed for the first time by Kittel in 1951 
\cite{Kittel.PhysRev.83.1951}. In the following decade, the existence of 
extrinsic contributions has been discussed several times in the literature, 
mainly in the context of Kittel's theory on the resonance character of 
permittivity of polydomain ferroelectric in the microwave frequency region due 
to effective domain wall inertia 
\cite{fatuzzo.JApplPhys.32.1965,Stanford.PhysRev.124.1961}. 
In 1960, Misařová \cite{Misarova.SovPhysSolState.2.1960} made an early 
discovery of putting the existence of extrinsic contributions to permittivity 
in connection with ferroelectric ageing. Nevertheless, a definite answer was 
not given to the question of the magnitude of the extrinsic contribution with 
respect to the intrinsic one until the work by Fousek 
\cite{Fousek.CzechJPhys.15.1965} where the methodology of the experimental 
determination of the extrinsic and intrinsic contributions has been 
established in a sense understood today, i.e. the intrinsic contribution 
corresponds to the dielectric response of the poled single-domain sample and 
the extra enhancement in the dielectric response, which is measured on a 
polydomain sample, corresponds to the extrinsic contribution. After this 
breakthrough in understanding of the concept, the extrinsic contributions to 
permittivity, piezoelectric coefficients, and elastic compliance became the 
subject of an intense theoretical and experimental research
\cite{Arlt.JApplPhys.58.1985,Robels.JApplPhys.73.1993,Xu.JApplPhys.89.2001,Taylor.JApplPhys.82.1997}.
Up to date, the extrinsic contributions have been considered something like a 
material curiosity of polydomain samples, which might be, however, 
beneficially used in applications. Nevertheless, the substantial drawback of 
the application of polydomain ferroelectric in electronic devices is the 
essential instability of the domain pattern, which is often manifested by a 
severe ageing of dielectric properties.

In this work, we present a method based on the comprehensive analysis of the 
evolution of the nonlinear dielectric response, which is caused by an 
evolution in the domain pattern microstructure. In Sec.~\ref{sec:II} we 
analyze the intrinsic and extrinsic contributions to the nonlinear 
dielectric response of the ferroelectric polydomain film in detail. 
Section~\ref{sec:III} presents a general method to identify the dominant 
microstructural phenomenon, which is responsible for the ageing of the 
ferroelectric. In Sec.~\ref{sec:IV} we demonstrate the application of our 
method on two qualitatively different systems. First, the system where the 
extrinsic permittivity is controlled by the bending movements of pinned domain 
walls (Sec.~\ref{sec:IVa}) and, second, the system with electrode-adjacent 
passive layers (Sec.~\ref{sec:IVb}). We show that both systems under 
consideration may manifest several microstructurally different ageing 
scenarios, which are characterized by different features of the evolution of 
nonlinear permittivity.

\section{Nonlinear dielectric response of the polydomain ferroelectric}
\label{sec:II}

\begin{figure}[t]
 \includegraphics[width=85mm]{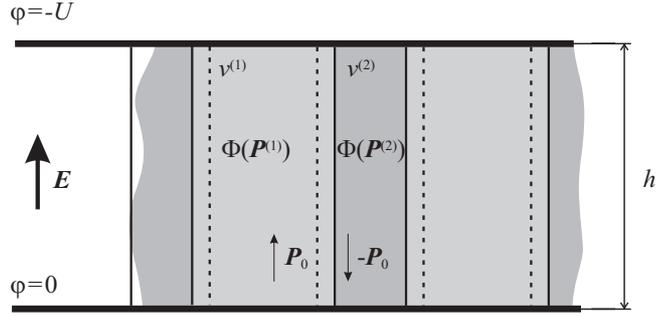}
\caption{
Scheme of the ferroelectric capacitor of thickness $h$ with the electric field 
$E$ applied to the material, which is in the polydomain state. Symbols $\1{v}$ 
and $\2{v}$ stand for the volume fractions of the ferroelectric, which is in 
the polarization state $\1{P}$ and $\2{P}$, respectively.
\label{fig:01}
}
\end{figure}
%
%
At first we analyze in detail the intrinsic and extrinsic contribution to the 
nonlinear dielectric response of the polydomain ferroelectric film, which is 
shown in Fig.~\ref{fig:01}. We consider a ferroelectric capacitor of thickness 
$h$ with the electric field $E$ applied to the material, which is split by 
180${}^\circ$ domain walls into the lamellar pattern of antiparallel domains 
at the two states with polarizations denoted by the symbols $\1{P_i}$ and 
$\2{P_i}$. The volume fractions of the domains are denoted by the symbols 
$\1{v}$ and $\2{v}$. In order to calculate the polarization response of the 
crystal lattice of the ferroelectric film, we adopt the incomplete 
thermodynamic potential (per unit volume of the sample) $g(P_i,\, E_i)
=\Phi(P_i)-P_iE_i$, which includes, first, the energy associated with the 
polarization of the crystal lattice $\Phi(P_i)$, and, second, the subtracted 
work produced by external voltage sources $-P_iE_i$. In order to present the 
basic principles of our method, we limit our considerations to the simplest 
case of the uniaxial ferroelectric with the vectors of the electric field 
$E_i$ and the polarization $P_i$ perpendicular to the top and bottom 
electrodes of the capacitor. A generalization to a more complicated system is 
more or less straightforward. For the simplification of the notation, we will 
omit the indices ``$i$" in the symbols for the vectors $P_i$ and $E_i$ in the 
following text. Then, the three lowest terms in the expansion of the function 
$g(P,E)$ with respect to the polarization $P$ of the crystal lattice equals:
\begin{equation}
	g(P,\,E)=\Phi_0 + \frac 12 \alpha P^2 + \frac 14 \beta P^4 - PE.
	\label{eq:01:Glat} 
\end{equation}
The polarization response of the crystal lattice can be found from the 
condition for the minimum of the thermodynamic potential $g$:
\begin{equation}
	\partial g(P,\,E)/\partial P=0.
	\label{eq:02:dGdPcond} 
\end{equation}
In the ferroelectric phase, when $\alpha<0$ and $\beta>0$, the polarization 
response in the two stable polarization states of the crystal lattice in the 
adjacent domains can be expressed in a form of Taylor series with respect to 
the electric field $E$:
\begin{subequations}
\label{eq:03:P}
\begin{eqnarray}
    \label{eq:03:P1}
    \1{P}(E)&\approx& 
			P_0 + 
			\varepsilon_0 \chi_c E - 
			\frac {3\varepsilon_0^2\chi_c^2}{2P_0}\, E^2 + 
			\frac {4\varepsilon_0^3\chi_c^3}{P_0^2}\, E^3 - \cdots,\\
    \label{eq:03:P2}
    \2{P}(E)&\approx& 
			- P_0 + 
			\varepsilon_0 \chi_c E + 
			\frac {3\varepsilon_0^2\chi_c^2}{2P_0}\, E^2 + 
			\frac {4\varepsilon_0^3\chi_c^3}{P_0^2}\, E^3 + \cdots,
\end{eqnarray}
\end{subequations}
where $P_0=\sqrt{-\beta/\alpha}$ is the spontaneous polarization and 
$\chi_c=-1/(2\varepsilon_0\alpha)$ is the dielectric susceptibility of the 
crystal lattice.
 
Now we can consider that the average polarization of the polydomain 
ferroelectric film $P_f$ is given by the volume weighted average of the 
polarization response in each domain state:
\begin{equation}
	P_f(E) = \1{v}\, \1{P}(E) + \2{v}\,\2{P}(E),
	\label{eq:04:Pavrg} 
\end{equation}
where $\1{v}+\2{v}=1$. Substituting Eqs.~(\ref{eq:03:P}) into Eq.~(\ref{eq:04:Pavrg}), one gets:
\begin{equation}
	P_f(E) = 
		\left(\1{v}-\2{v}\right)\,
		\left(P_0 - \frac {3\varepsilon_0^2\chi_c^2}{2P_0}\, E^2\right) +  
		\varepsilon_0 \chi_c E + 
		\frac {4\varepsilon_0^3\chi_c^3}{P_0^2}\, E^3 + \cdots.
	\label{eq:05:PavrgIs} 
\end{equation}
It is convenient to express the above expression in terms of the net 
spontaneous polarization $P_N = (\1{v}-\2{v})\, P_0$:
\begin{equation}
	P_f(E) = 
		P_N(E)\,
		\left(1 - \frac {3\varepsilon_0^2\chi_c^2}{2P_0^2}\, E^2\right) +  
		\varepsilon_0 \chi_c E + 
		\frac {4\varepsilon_0^3\chi_c^3}{P_0^2}\, E^3 + \cdots.
	\label{eq:06:PavrgIs2} 
\end{equation}
The function $P_N(E)$ has a meaning of the extrinsic polarization response of 
the polydomain film to the electric field; i.e. the response, which is 
produced by the change in volume fractions of the two domain states due to the 
domain wall motions.

The response of the net spontaneous polarization to the electric field $P_N(E)
$ is controlled by nonlocal macroscopic phenomena, such as interaction of the 
domain walls with crystal lattice defects
\cite{Tagantsev.Ferroelectrics.221.1999,Mokry.ISAF16.2007}, 
depolarizing fields due to the presence of electrode-adjacent layers 
\cite{Kopal.Ferroelectrics.223.1999,Bratkovsky.PhysRevB.63.2001,Mokry.PhysRevB.70.2004}, 
reduction of the domain wall mobility due to free charges 
\cite{Mokry.Ferroelectrics.319.2005,Mokry.PhysRevB.75.2007},
etc. In order to calculate the function $P_N(E)$ we adopt the thermodynamic 
potential $G_w(P_N,E)$ (per unit area of the ferroelectric capacitor) of the 
system with the constant electric field $E$. The function $G_w(P_N,E)$ 
includes two contributions: First, the free energy, which is increased when 
the domain wall is shifted from its equilibrium position (e.g. energy of the 
depolarizing field, surface energy of the domain walls, etc.), and, the 
subtracted work produced by the external sources that keeps the constant 
voltage $U=Eh$ on the electrodes (see Fig.~\ref{fig:01}). In the case of the 
non-poled (or depolarized) polydomain ferroelectric sample, the three lowest 
terms in the expansion of the function $G_w(P_N,E)$ with respect to the net 
spontaneous polarization $P_N$ equal:
\begin{equation}
	G_w(P,\,E)= 
		h\,\left[
			\frac 12 \alpha_w P_N^2 + \frac 14 \beta_w P_N^4 - P_NE
		\right].
	\label{eq:07:Gw} 
\end{equation}
In the case of the stable domain pattern, when $\alpha_w>0$, the extrinsic 
polarization response is given by the condition for the minimum of the 
function $G_w(P_N,E)$:
\begin{equation}
	\partial G_w(P_N,\,E)/\partial P_N=0
	\label{eq:08:dGwdPNcond} 
\end{equation}
and the net spontaneous polarization $P_N$ can be expressed in the form of 
the Taylor series with respect to the electric field $E$:
\begin{equation}
    \label{eq:09:PN}
    P_N(E) = 
			\varepsilon_0 \chi_w E + 
			\frac {1}{3}\varepsilon_0 b_w\, E^3 + \cdots,
\end{equation}
where $\chi_w=1/(\varepsilon_0\alpha_w)$ is the extrinsic contribution to  
dielectric susceptibility of the system and 
$b_w=-3\beta_w/(\varepsilon_0\alpha_w^4)$ is the nonlinearity constant of the 
extrinsic contribution to dielectric susceptibility of the system with 
respect to the electric field.

Substituting Eq.~(\ref{eq:09:PN}) into Eq.~(\ref{eq:06:PavrgIs2}) one gets 
the average nonlinear polarization response of the polydomain ferroelectric in 
the form:
\begin{equation}
	P_f(E) =
		\varepsilon_0 \left(\chi_c + \chi_w\right)\, E +
		\frac 13\, \varepsilon_0 \left(b_c + b_w\right) \, E^3 + \cdots,
	\label{eq:10:PavrgIs3} 
\end{equation}
where
\begin{equation}
	b_c = \frac{12 \, \varepsilon_0^2\chi_c^3}{P_0^2}
		\left(
			1 - \frac{3\chi_w}{8\chi_c}
		\right)
	\label{eq:11:bc} 
\end{equation}
is the nonlinearity constant due to the nonlinear dielectric response of the 
crystal lattice. It should be noted that the value of $b_c$ is also controlled 
by the linear extrinsic contribution to permittivity $\chi_w$. Finally, the 
effective dielectric susceptibility of the polydomain ferroelectric layer 
$\chi_f=(1/\varepsilon_0)(\partial P_f/\partial E)$ is equal to: 
\begin{equation}
	\chi_f(E) \approx
		\left(\chi_c + \chi_w\right)+
		\left(b_c + b_w\right) \, E^2.
	\label{eq:12:epsf} 
\end{equation}

\section{Method to study ageing of the polydomain ferroelectric}
\label{sec:III}

As it has already been mentioned, we consider that the extrinsic contribution to 
the nonlinear dielectric response $\chi_w + b_w\,E^2$ is controlled by a 
particular nonlocal phenomenon associated with some microstructural mechanisms 
affecting the quality of the ferroelectric. It is reasonable to consider that in 
the systems with the pronounced ageing of ferroelectric properties, the 
associated microstructural mechanism should be also responsible for the 
evolution of the parameters $\chi_w$ and $b_w$. Therefore, it is reasonable to 
expect that there exists a specific relation between the parameters $\chi_w$ 
and $b_w$ during ageing, which can be expressed in a form $\chi_w=F(b_w)$. The 
crucial point of the presented method is the fact, that the explicit form of 
the function $F$ is unique for the every particular microstructural mechanism 
of the dielectric ageing. Hence, when the experimentally observed dependence of 
$F(b_w)$ vs. $\chi_w$ matches the theoretical prediction, there exists a 
reasonable indication that the considered mechanism is responsible for the 
observed ageing of the dielectric response. It means that this method can be 
used for the identification of the origin of the ageing of ferroelectric 
films. In this section, we show a general way to determine the 
characteristic function $F$.

At first, we shall start with the fact that in the most of the ferroelectric 
materials with pronounced ageing of the dielectric properties, the dielectric 
nonlinearity is dominated by the contribution of the domain wall motion, 
i.e. $b_c\ll b_w$. Then we consider that the evolution of the dielectric 
response is induced by the evolution in some ``internal" microstructural 
parameter (e.g. average density of the pinning centers on the domain wall, 
average domain wall spacing, thickness of the electrode-adjacent dead layer, 
spontaneous polarization screening by free charges, etc.), which we denote by 
a general symbol $\xi$. We consider that this parameter controls the values of 
$\chi_w$ and $b_w$ and we can write:
\begin{subequations}
\label{eq:13:fg}
\begin{eqnarray}
    \label{eq:13:fga}
    \chi_w &=& f(\xi),\\
    \label{eq:13:fgb}
    b_w &=& g(\xi),
\end{eqnarray}
\end{subequations}
and hence    
\begin{equation}
	\chi_f(E) \approx
		\chi_c + f(\xi) + g(\xi) \, E^2.
	\label{eq:14:epsf2} 
\end{equation}
If the function $g(\xi)$ is monotonic, we can express the parameter $\xi$ from 
Eq.~(\ref{eq:13:fgb}), i.e. $\xi=g^{-1}(b_w)$ and substitute it into 
Eq.~(\ref{eq:13:fga}):
\begin{equation}
	\chi_w = f\left[g^{-1}\left(b_w\right)\right] = F\left(b_w\right).
	\label{eq:15:h} 
\end{equation}
Now, if we consider that the intrinsic contribution to the dielectric 
susceptibility $\chi_c$ does not change in time, the function $F$ can be 
cross-checked by the analysis of the data extracted from the dielectric ageing 
measurements on the polydomain ferroelectric films. The first measured 
parameter is the small signal permittivity $\varepsilon_L$:
\begin{equation}
	\varepsilon_L = 1+\chi_f(0) = \varepsilon_c + \chi_w,
	\label{eq:16:epsL} 
\end{equation}
where $\varepsilon_c=1+\chi_c$ is the relative permittivity of the crystal 
lattice. The second measured parameter is the dielectric nonlinearity constant 
 $b$, which can be determined from the following expression:
\begin{equation}
	b\, E^2 = \varepsilon_f(E) - \varepsilon_L,
	\label{eq:17:bM} 
\end{equation}
where $\varepsilon_f(E)$ is the electric field dependence of the permittivity 
of the polydomain ferroelectric film. 

To apply our method on the experimental data, one should take two steps: first,
 to prove the quadratic dependence of the permittivity on the applied electric 
field given by Eq. (\ref{eq:12:epsf}) and, second, to identify the 
relationship between the small signal extrinsic permittivity $\chi_w$ and the 
dielectric nonlinearity parameter $b$ given by Eq. (\ref{eq:17:bM}) during the 
dielectric ageing. Actually, verifying the first step represents the 
experimental evidence for a fast reversible movement of the domain walls, 
which should be distinguished from the Rayleigh-type relation, where the 
dielectric permittivity increases linearly with the ac-field as a result of 
the irreversible movement of domain walls under the sub-switching fields 
(usually a few tens of kV/cm)
\cite{Taylor.ApplPhysLett.73.1998,Xu.JApplPhys.89.2001,Zhang.JApplPhys.100.2006}. 
In the second step, we can identify the particular ageing scenario from the 
experimental evidence of the mutual relationship between the parameters 
 $\chi_w$ and $b_w$ in time during the dielectric ageing. By combining Eqs. 
(\ref{eq:15:h}) and (\ref{eq:16:epsL}) one gets
\begin{equation}
	F(b) = \chi_w = \varepsilon_L - \varepsilon_c.
	\label{eq:18:test} 
\end{equation}
Therefore the validity of Eq. (\ref{eq:18:test}) can be demonstrated by the 
linear relationship between the values of $F(b)$ and the total small field 
dielectric permittivity $\varepsilon_L$. The relationship given in Eq. 
(\ref{eq:18:test}) can be tested experimentally by fitting the values of 
$F(b)$ and $\varepsilon_L$ to the function of the form: 
\begin{equation}
	F(b) = K\, \varepsilon_L - B.
	\label{eq:18:fit} 
\end{equation}
If the considered microstructural mechanism is responsible for the evolution 
of the dielectric response, the fitted value of the parameter $K$ should be 
approximately equal to unity, i.e. $K\approx 1$, and the fraction $B/K$ 
should correspond to the value of crystal lattice permittivity, i.e. 
$\varepsilon_c\approx B/K$, which can be checked by the measurements on the 
polarized single-domain sample.

\section{Evolution of the nonlinear dielectric response during ageing}
\label{sec:IV}

In this section, we demonstrate the applicability of the general method 
presented above for two particular systems. At first we present the system, 
where the extrinsic permittivity is controlled by bending movements of the 
pinned domain walls. Then we demonstrate the system with electrode-adjacent 
passive layers, where the domain wall movements are controlled by the 
depolarizing field in the electrode-adjacent passive layer.

\subsection{Ferroelectric ageing in the system with pinned domain walls}
\label{sec:IVa}

In this subsection, we adopt a model presented in \cite{Mokry.ISAF16.2007}, 
where the extrinsic permittivity of the polydomain ferroelectric is controlled 
by bending movements of the pinned 180${}^\circ$ domain walls. This scenario seems 
to be quite reasonable due to several direct observations of such interactions 
of the domain walls with bulk crystal lattice defects 
\cite{Yang.PhysRevLett.82.1999}. In addition, recent ab-initio calculations 
\cite{He.PhysRevB.68.2003,Meyer.PhysRevB.65.2002} have indicated that the 
oxygen vacancy, which is a very common defect in the perovskite ferroelectrics, 
has a smaller formation energy in the 180${}^\circ$ domain wall than in the 
bulk and, thus, the oxygen vacancy accompanied with two free electrons to 
comply the requirement of the electroneutrality of the system can be 
responsible for the pinning effect on 180${}^\circ$ domain walls.

\begin{figure}[t]
 \includegraphics[width=85mm]{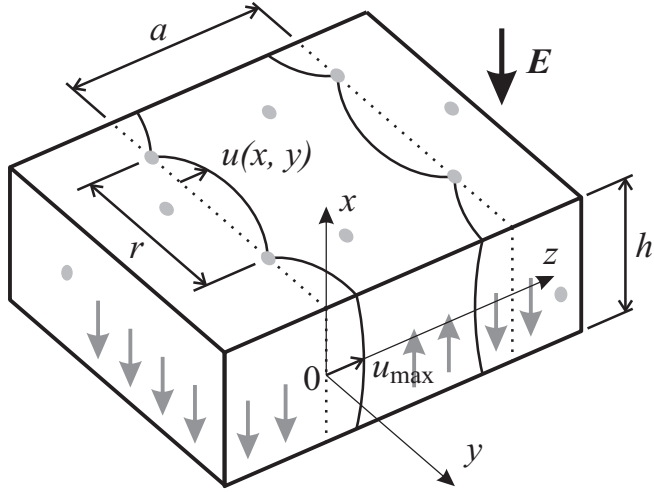}
\caption{
Model of the 180${}^\circ$ domain wall bending. Planar position of the domain 
wall (doted line) is locally blocked by the pinning centers (gray circles). 
Orientation of vector of spontaneous polarization is indicated by gray arrows. 
When the external electric field $E$ is applied to the ferroelectric film, the 
domain wall is bent between the pinning centers and the profile of the wall 
deflection is described by the function $u(x, y)$.
\label{fig:02}
}
\end{figure}
%
%
Figure~\ref{fig:02} shows the model of bending movements of pinned 
180${}^\circ$ ferroelectric domain wall with the attached Cartesian coordinate 
system. We consider that the pinning centers on the domain wall are 
distributed in the $x$ and $y$ directions with an average distance $r$ between 
them and, thus, the pinning centers density on the domain wall is proportional 
to $1/r^{2}$. When the electric field $E$ is applied to the ferroelectric film 
along the ferroelectric axis, the originally planar domain wall is bent 
between pinning centers, as indicated in Fig.~\ref{fig:02}, and the profile of 
the wall deflection is described by a function $u(x,\,y)$. The domain wall 
spacing is denoted by a symbol $a$.

When the domain wall is bent, the thermodynamic function $G_w$ consists of two 
contributions: first, the surface energy of the bent domain wall and, second, 
the electrostatic energy associated with the depolarizing field due to 
discontinuous change of the normal component of the spontaneous polarization at 
the domain wall. At the constant electric field $E$, the thermodynamic 
potential $G_w$ is given by the formula:
\begin{equation}
	\label{eq:19:GwdefPIN} 
 	G_w
	=
	\frac{1}{ar^2}\left\{ -2P_0 E\int_A u(x,y)dA +
	\int_A
	\left[
		\sigma _w + \frac 12 \sigma _b \varphi _d
	\right]
	\sqrt {1+u_x^2 (x,y)+u_y^2 (x,y)} dA  \right\}
\end{equation}
where the integrals are taken over the square $A$ where $x$ and $y$ are 
running from $-r/2$ to $r/2$, a symbol $\sigma_{w}$ stands for the surface 
energy density associated with the surface tension of the bent domain wall, 
and a symbol $\varphi_{d}$ stands for the electrostatic potential on the 
domain wall associated with the depolarizing field, which is produced by the 
bound charges of the surface density $\sigma_{b}$ due to the discontinuous 
change of the normal component of the spontaneous polarization at the domain 
wall. The functions $u_{x}(x,y) $ and $u_{y}(x,y)$ are the partial derivatives 
of the function $u(x,y)$ with respect to $x$ and $y$, respectively. The 
function $\varphi_{d}$ can be readily calculated by solving the Poisson 
equation for the electrostatic potential and by considering the conditions for 
the continuity of the normal component of the electric displacement and for 
the continuity of the electrostatic potential. If one considers that the net 
spontaneous polarization $P_{N}$ is equal to:
\begin{equation}
	\label{eq:20:PNisPIN} 
	P_N =\frac{2P_0 }{ar^2}\int_A {u(x,y)\,dA},
\end{equation}
and that the maximum deflection $u_{max}$ of the domain wall is much smaller 
than the average distance between the pinning centers $r$, our detailed 
calculations \cite{Mokry.PhysRevB.bending.unpubl1} show that the leading terms 
in the Taylor expansion of the thermodynamic function $G_w$ in this particular 
system with respect to the net spontaneous polarization are:
\begin{equation}
	\label{eq:21:GwPIN} 
	G_w(P_N,\, E) \approx 
		h\,\left[
			\frac{a\, a_w }{8\varepsilon _0 \varepsilon_c r^2}\, P_N^2 -  
			\frac{0.17\,a^3a_w}{\varepsilon _0 \varepsilon_c r^4P_0 }\, P_N^4 - 
			P_N E
		\right],
\end{equation}
where the symbol $a_w$ stands for the thickness of the domain wall. 
Solving Eq.~(\ref{eq:08:dGwdPNcond}) for this particular system, one gets:
\begin{subequations}
\label{eq:22:chiwbwPIN}
\begin{eqnarray}
    \label{eq:22:chiwbwPINa}
    \chi_w(r,\,a) &\approx& 
			\frac{4\varepsilon _c r^2}{a\,a_w },\\
    \label{eq:22:chiwbwPINb}
    b_w(r,\,a) &\approx& 
			\frac{518.4\,r^4\varepsilon_0^2 \varepsilon_c^3 }{a\, a_w^3 P_0^2}.
\end{eqnarray}
\end{subequations}

Now in this particular system, we can distinguish two scenarios of the 
dielectric ageing.

\subsubsection{Dielectric ageing due to the progressive pinning}

\noindent
In this scenario, we consider, that due to the drift of the crystal lattice 
defects (e.g. oxygen vacancies) in the material, the concentration of the 
pinning centers on the domain wall may be increasing during the dielectric 
ageing. Therefore, in this scenario, which is characterized by a progressive 
pinning of the domain walls, we consider that the microstructural parameter 
responsible for the dielectric ageing is the average distance $r$ of the 
pinning centers and that the domain wall spacing $a$ remains constant.

Then we can express the parameter $r$ from Eq.~(\ref{eq:22:chiwbwPINb}) and by 
substituting it into Eq.~(\ref{eq:22:chiwbwPINa}) we arrive at the following 
form of the function $F(b)$:
\begin{equation}
	\label{eq:23:hPIN} 
	F(b) =
		\sqrt{b}\, 
		\sqrt{
			\frac{0.031\, P_0^2}{\varepsilon_0^2\varepsilon_c}
			\left(
				\frac{a_w}{a}
			\right)
		}. 		
\end{equation}
In addition, from the time evolution of the parameters $b(t)$ and 
$\varepsilon_L(t)$ we can estimate the evolution of the average distance $r(t)
$ between pinning centers on the domain wall:
\begin{equation}
	\label{eq:24:rEvolPIN} 
	r(t) =
		\frac{
			0.088\, a_w P_0
		}{
			\varepsilon_0 \varepsilon_c
		}
		\sqrt{
			\frac{b(t)}{\varepsilon_L(t)-\varepsilon_c}
		}. 		
\end{equation}

\subsubsection{Dielectric ageing due to the domain wall coalescence}

\noindent
On the other hand, the same microstructural model offers a different ageing 
scenario. In the ferroelectric polydomain film, it is possible that the domain 
pattern is in the essentially non-equilibrium configuration and, therefore, 
there exists a ``thermodynamic" force applied to the domain wall that drives 
the system to reach the ``absolute" equilibrium configuration. In this 
situation, the depinning of some walls can occur and they can coalesce to 
reduce the energy proportional to the domain wall area. Then, we can consider 
that the parameter responsible for the evolution of the dielectric response is 
the domain spacing $a$ and that the concentration of the pinning centers on 
the domain wall remains constant during ageing. Hence, we can express the 
parameter $a$ from Eq.~(\ref{eq:22:chiwbwPINb}) and by substituting it into 
Eq.~(\ref{eq:22:chiwbwPINa}) we obtain the characteristic function $F(b)$ in 
the form:
\begin{equation}
	\label{eq:25:hPINa} 
	F(b) =
		b\,
		\left[
			\frac{0.088\, P_0}{\varepsilon_0\varepsilon_c}
			\left(
				\frac{a_w}{r}
			\right)
		\right]^2. 		
\end{equation}
Similarly as in the previous ageing scenario, it is possible to estimate the 
time evolution of the average domain spacing $a(t)$ from the time evolution of 
the parameters $b(t)$ and $\varepsilon_L(t)$:
\begin{equation}
	\label{eq:26:aEvolPIN} 
	a(t) =
		\frac{
			0.031\, a_w P_0^2\, b(t)
		}{
			\varepsilon_0^2 \varepsilon_c\, \left[\varepsilon_L(t)-\varepsilon_c\right]^2
		}.
\end{equation}
  
Thus, it is seen that using cross-checking the two qualitatively different $b$ 
vs. $\varepsilon_L$ dependences, it is possible to distinguish the 
microstructural processes, which are responsible for the evolution of the 
nonlinear dielectric response.

\subsection{Ferroelectric ageing in the system with the electrode-adjacent passive layer}
\label{sec:IVb}

\begin{figure}[t]
 \includegraphics[width=85mm]{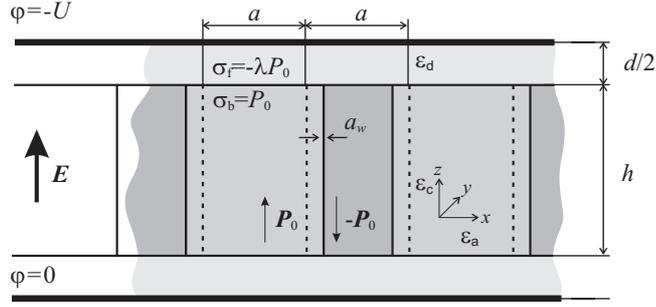}
\caption{
Scheme of the electroded ferroelectric film with passive layers. It is always 
considered that $h\gg d$.
\label{fig:03}
}
\end{figure}
%
%
In this subsection, we study the dielectric response of the system, where the 
ferroelectric layer with the antiparallel domain pattern is separated from the 
electrodes by the electrode-adjacent passive layers. Although the nature of these 
layers is not clear \cite{Sinnamon.ApplPhysLett.78.2001} and the direct 
observations of well-defined layers are rather scarce 
\cite{Hase.JpnJApplPhys.32.1993}, numerous indirect experimental observations 
support the wide acceptance of the passive layer concept. Until today, 
various mechanisms of film-electrode interactions have been proposed: creation 
of variable depletion layers associated with the Schottky barriers 
\cite{Dey.JpnJApplPhys.34.1995,Basceri.JApplPhys.82.1997}, low permittivity 
layers due to the oxygen vacancy entrapment at ferroelectric-electrode 
interfaces \cite{Lee.JApplPhys.78.1995}, local suppression of polarization 
states in the near-electrode regions \cite{Vendik.JApplPhys.88.2000}, local 
diffusion of the electrode material into the ferroelectric 
\cite{Stolichnov.ApplPhysLett.75.1999}, or a chemically distinct surface phase 
\cite{Craciun.ApplPhysLett.76.2000}.
 
In order to calculate the dielectric response of the system with a passive 
layer, we adopt the model shown in Fig.~\ref{fig:03}. We consider a sandwich 
structure of the ferroelectric layer of the thickness $h$ with a lamellar 
antiparallel domain pattern, two dielectric passive layers of the total 
thickness $d$ and the permittivity $\varepsilon_d$, which separate the 
ferroelectric layer from the electrodes. The spontaneous polarization $P_0$ is 
considered to be constant within each domain and its abrupt change to zero at 
the interface of the ferroelectric and dielectric layers ($z=\pm h/2$) leads 
to the appearance of a bound charge of the surface density $\sigma_b=P_0$. Due 
to finite conductivity of the passive layer, the bound charge is considered to 
be compensated by the free charge injected from the electrodes, so the 
proportionality factor $\lambda$ called the \emph{degree of screening} is 
introduced to measure the process of the charge transport 
 $\sigma_f=-\lambda\sigma_b$, where $0\leq\lambda\leq 1$.

In this model, we neglect the effect of the lattice pinning of the domain wall 
and we consider, for simplicity, that the net spontaneous polarization 
response to the electric field is controlled by the electrostatic energy of 
the depolarizing field produced by the periodic distribution of the bound 
 $\sigma_b$ and free $\sigma_f$ charges at the interface of the ferroelectric 
and dead layers. In addition, we address in detail the case of the neutral 
domain pattern, i.e. we set $P_N=0$ in the absence of the applied voltage 
 $U=0$. The thermodynamic potential $G_w$ (per unit area of the sample) of a 
system with a given voltage $U$ on the electrodes and free charges $\sigma_f$ 
inside the system, which includes the electrostatic energy of the electric 
field with the subtracted work produced by the external voltage sources, is 
given by the formula:
\begin{equation}
    G_w = \frac1{2A} \int_V E_i\left(D_i-P_{0,i}\right)\,dV +
    U\sigma_0,
	\label{eq:27:GwdefSCR} 
\end{equation}
where $A$ is the area of the ferroelectric capacitor and $\sigma_0$ is the 
surface density of free charges on the electrodes. 

Calculation of the electrostatic energy requires the knowledge of the 
electrostatic potential $\varphi$, which satisfies the following equations of 
electrostatics in the ferroelectric and the passive layers,
\begin{equation}
    \varepsilon_a\frac{\partial^2\varphi_f}{\partial x^2} +
    \varepsilon_c\frac{\partial^2\varphi_f}{\partial z^2} = 0,
    \quad
    \frac{\partial^2\varphi_g}{\partial x^2} +
    \frac{\partial^2\varphi_g}{\partial z^2} = 0,
    \label{eq:26:EqLap}
\end{equation}
with the boundary conditions $\varphi=-(+)U/2$ at $z=+(-)(h+d)/2$, and
\begin{equation}
    \varepsilon_c\frac{\partial\varphi_f}{\partial z} -
    \varepsilon_d\frac{\partial\varphi_g}{\partial z} =
    \frac{\sigma(x)}{\varepsilon_0},
    \quad
    \varphi_f=\varphi_d,
    \quad
    \mbox{at } z=h/2,
    \label{eq:27:PhiCond}
\end{equation}
where the subscripts $f$ and $g$ denote the ferroelectric and passive layer, 
respectively. Here $\sigma(x)=\sigma_b(x)+\sigma_f(x)$ is the sum of the 
surface density of the bound charge due to the spontaneous polarization and 
the free screening charge from the electrode. Solution of the potential and 
the calculation of the electrostatic energy for a similar model are already 
available in the literature
\cite{Kopal.Ferroelectrics.223.1999,Bratkovsky.PhysRevB.63.2001,Mokry.PhysRevB.70.2004}, 
however with a different distribution of charges at the interface of the 
ferroelectric and passive layers. In this model, we consider that the 
spontaneous polarization continuously and smoothly changes its value within a 
distance $a_w$ representing the domain wall thickness. Then the function 
$\sigma(x)$ can be obtained in a form of the Fourier series:
\begin{equation}
    \sigma(x)
		=
		P_N +
        \sum\limits_{n=1}^\infty \frac{64a^2P_0}{a_w^2\pi^3n^3}\,
            \sin^2\left(\frac{n\pi a_w}{4a}\right)
            \cos\left(\frac{n\pi x}{a}\right)
            \left\{
                \sin\left[
                    \frac{n\pi}2
                    \left(
                        1 + \frac{P_N}{P_0}
                    \right)
                \right]
                    -\lambda
                \sin\left(
                    \frac{n\pi}2
                \right)
            \right\}.
    \label{eq:28:sigma}
\end{equation}

In order to calculate the dielectric response, we express the function $G_w$ 
in terms of the applied electrode voltage $U$ and the net spontaneous 
polarization of the ferroelectric layer $P_N$. With use of the results 
published earlier
\cite{Kopal.Ferroelectrics.223.1999,Bratkovsky.PhysRevB.63.2001,Mokry.PhysRevB.70.2004}, 
the leading terms of the function $G_w$ in this model are equal to:
\begin{equation}
    G_w(P_N,\, E)=h\left[
        	\frac12\,\alpha_w P_N^2 + 
			\frac14\,\beta_w P_N^4 -
			P_N E
    \right],
    \label{eq:29:GwSCR}
\end{equation}
where
\begin{subequations}
\label{eq:30:constantsSCR}
\begin{eqnarray}
    \alpha_w(\lambda,\,d) &=&
        	\frac d{\varepsilon_0\left(\varepsilon_d h + \varepsilon_c d\right)}
			+
			\frac{1.67 a^5}{\varepsilon_0 a_w^4 h}\,
			\sum\limits_{n=1}^\infty
            \frac{\lambda -
            (-1)^n(\lambda-2)}{n^5D_n(d/a)}\,
				\sin^4\left(\frac{n\pi a_w}{4a}\right),
    \label{eq:30:alphawSCR}	\\
    \beta_w(\lambda,\,d) &=&
        \frac{0.688\, a^5}{\varepsilon_0 a_w^4 h P_0^2}\,
        \sum\limits_{n=1}^\infty
            \frac{(-1)^n(\lambda-8) - \lambda}{n^3D_n(d/a)}\,
				\sin^4\left(\frac{n\pi a_w}{4a}\right),
    \label{eq:30:betawSCR}	\\
    E &=& \frac{\varepsilon_d U}{\varepsilon_d h + \varepsilon_c d},
    \label{eq:30:EwSCR}	
\end{eqnarray}
and
\begin{equation}
    D_n(\tau) = \varepsilon_d\, \coth\frac{n\pi\tau}{2} +
        \sqrt{\varepsilon_a\varepsilon_c}\,
        \coth\left(\sqrt{\frac{\varepsilon_a}{\varepsilon_c}}\frac{n\pi h}{2a}\right).
    \label{eq:30:Dn}	
\end{equation}
\end{subequations}

Similarly as in the previous section also in this particular system we can 
distinguish two ageing scenarios:

\subsubsection{Dielectric ageing due to the thickening of the electrode-adjacent passive layer}

It was reported \cite{Lee.JApplPhys.78.1995,Sun.ApplPhysLett.80.2002} that the 
fatigue of the dielectric response of the ferroelectric films can be caused by 
an increase of the thickness of the passive layer (e.g. due to the oxygen 
vacancy entrapment at the film-electrode interface). The main indication, from 
which such a statement is inferred, was the measurement of the size effect on 
the permittivity. There there was observed a decrease of the effective 
capacitance of the interface layer. Unfortunately, it was also demonstrated 
\cite{Mokry.Ferroelectrics.319.2005} that the same effect on the permittivity, 
i.e. the effective decrease of the interface layer capacitance, can be 
produced by the spontaneous polarization screening by immobile charged 
defects. Nevertheless, by applying the present method on the data of the 
nonlinear permittivity, it was possible to distinguish whether such a fatigue 
scenario may take place.

In order to make it possible to use the principles of the discussed method to 
this particular situation, we should make following considerations: (i) The 
dominant phenomenon, which controls the extrinsic dielectric response, is the 
increase of the passive layer thickness, $d>0$; (ii) There is no spontaneous 
polarization screening, $\lambda=0$; (iii) The domain pattern in the 
ferroelectric layer is dense, $a\ll h$; (iv) The passive layer in the system 
is thin compared to the average domain spacing, $d\ll a$; (v) The thickness of 
the domain wall is much smaller than the domain spacing, $a_w\ll a$; (vi) The 
permittivity of the passive layer can be approximated by the permittivity of 
the ferroelectric in the following way: 
$\varepsilon_d=\sqrt{\varepsilon_a\varepsilon_c}$; Finally (vii), the average 
domain spacing $a$ remains constant during the dielectric ageing. Under these 
considerations, Eqs.~(\ref{eq:30:constantsSCR}) can be further simplified and 
Eq.~(\ref{eq:29:GwSCR}) has a simple form:
\begin{equation}
    G_w(P_N,\, E)\approx h\left[
			\frac{
				d^2\pi
			}{
				8\, a h\, \varepsilon_0 \sqrt{\varepsilon_a\varepsilon_c}
			}\, 
			P_N^2 + 
			\frac{
				d^2\pi^3
			}{
				192\, a h\, \varepsilon_0 P_0^2 \sqrt{\varepsilon_a\varepsilon_c}
			}\, 
			P_N^4 -
			P_N E
    \right],
    \label{eq:34:GwSCRb}
\end{equation}
Solving Eq.~(\ref{eq:08:dGwdPNcond}) for this particular situation, one gets:
\begin{subequations}
\label{eq:35:chiwbwSCRb}
\begin{eqnarray}
    \chi_w(d) &\approx&
			\frac{
				4\, ah \sqrt{\varepsilon_a\varepsilon_c}
			}{
				d^2\, \pi
			},
	 \label{eq:35:chiwbwSCRba} \\
    b_w(d) &\approx&
        -\frac{
				16\,a^3 h^3 \varepsilon_0^2\left(\varepsilon_a\varepsilon_c\right)^{3/2}
			}{
				\pi\, d^6 P_0^2
			}.
	 \label{eq:35:chiwbwSCRbb}
\end{eqnarray}
\end{subequations}
Finally, the function $F(b)$ is obtained by expressing the parameter $d$ from 
Eq.~(\ref{eq:35:chiwbwSCRbb}) and by substituting it into 
Eq.~(\ref{eq:35:chiwbwSCRba}):
\begin{equation}
    F(b) = 
		\sqrt[3]{-b}\, 
		\sqrt[3]{
			\frac{0.41\, P_0^2}{\varepsilon_0^2}}.
    \label{eq:36:hSCRb}	
\end{equation}

\subsubsection{Dielectric ageing due to the spontaneous polarization screening}

Here we show that within the same microstructural model of the passive layer, 
we can discuss the possibility of a different ageing scenario. 
Due to the essentially semiconductive nature of ferroelectric materials and 
due to the strong experimental evidence 
\cite{Stolichnov.JApplPhys.84.1998,Tagantsev.JApplPhys.90.2001}, 
the free charge injection across the electrode-adjacent layer has become a 
widely accepted concept. Nevertheless, it has been discussed mainly in 
connection with the ferroelectric switching measurements and the phenomenon 
called imprint, i.e. the horizontal shift of the hysteresis loop, which is 
caused by the internal bias field produced by the injected free charges. In 
this subsection, we deal with the free charge injection across the 
electrode-adjacent layer in the system with the polydomain ferroelectric layer.

It has been already reported \cite{Sun.ApplPhysLett.80.2002} that the pinning 
of domain walls in the electrode-adjacent region may represent a source of the 
essential nonlinearity of the dielectric response. It has been also reported 
\cite{Mokry.Ferroelectrics.319.2005} that the similar pinning effect on the 
domain walls can be produced by the low mobility of the free charge at the 
film-electrode interface. Here we study how the injected free charges affect 
the nonlinear dielectric response of polydomain films.

In order to make a clear statement, whether the free charge injection across 
the passive layer can be responsible for the observed dielectric ageing using 
the presented method, we shall make the following considerations: (i) The 
dominant phenomenon, which controls the extrinsic dielectric response is the 
spontaneous polarization screening, $\lambda>0$; (ii) The domain pattern in 
the ferroelectric layer is dense, $a\ll h$; (iii) The passive layer in the 
system is thin compared to the average domain spacing, $d\ll a$; (iv) The 
thickness of the domain wall is much smaller than the domain spacing, $a_w\ll 
a$; (v) The permittivity of the passive layer can be approximated by the 
permittivity of the ferroelectric in a following way, 
$\varepsilon_d=\sqrt{\varepsilon_a\varepsilon_c}$; (vi) Since we are focused 
on the deteriorating impact of the free screening charges on the dielectric 
response of the polydomain ferroelectric film, which is produced due to the 
pinning effect of the free charges on the domain walls, we further consider that 
the free charges are of much lower mobility than the mobility of the domain 
wall in the absence of the free charges; Finally (vii), the average domain spacing 
$a$ remains constant during the dielectric ageing. Under these considerations, 
Eqs.~(\ref{eq:30:constantsSCR}) can be further simplified and 
Eq.~(\ref{eq:29:GwSCR}) has a simple form:
\begin{equation}
    G_w(P_N,\, E)\approx h\left[
			\frac{
				2\, \lambda\, a d
			}{
				3\, \varepsilon_0  a_w h \sqrt{\varepsilon_a\varepsilon_c}
			}\, 
			P_N^2 - 
			\frac{
				\lambda\, a^3 d
			}{
				6\, \varepsilon_0 a_w^3 h P_0^2 \sqrt{\varepsilon_a\varepsilon_c}
			}\, 
			P_N^4 -
			P_N E
    \right],
    \label{eq:31:GwSCR}
\end{equation}
where $E=U/h$. Solving Eq.~(\ref{eq:08:dGwdPNcond}) for this particular 
situation, one gets:
\begin{subequations}
\label{eq:32:chiwbwSCR}
\begin{eqnarray}
    \chi_w(\lambda) &\approx&
			\frac{
				3\, a_w h\, \sqrt{\varepsilon_a\varepsilon_c}
			}{
				4\, a d \lambda
			},
	 \label{eq:32:chiwbwSCRa} \\
    b_w(\lambda) &\approx&
        \frac{
				0.633\,a_w h^3 \varepsilon_0^2\left(\varepsilon_a\varepsilon_c\right)^{3/2}
			}{
				a d^3 \lambda^3 P_0^2
			}.
	 \label{eq:32:chiwbwSCRb}
\end{eqnarray}
\end{subequations}
It should be stressed again that the limit for $\lambda$ approaching zero 
cannot be applied to Eqs.~(\ref{eq:32:chiwbwSCR}), since it violates the 
conditions for the applicability of our approximations. The reason for it 
can be described as follows. When the degree of screening $\lambda$ is smaller 
than some critical value $\lambda_{\rm crit}$, the leading terms that control 
the thermodynamic function $G_w$ given by Eqs. (\ref{eq:29:GwSCR}) and 
(\ref{eq:30:constantsSCR}) become very weakly controlled by the spontaneous 
polarization screening $\lambda$ and the other parameters of the system have 
stronger effect on the net spontaneous response of the system. The critical 
value $\lambda_{\rm crit}$ can be easily calculated by comparing the quadratic 
term coefficients in the expansion of the thermodynamic function $G_w$ with 
respect to the net spontaneous polarization in the two aforementioned different 
approximations given by Eqs. (\ref{eq:34:GwSCRb}) and (\ref{eq:31:GwSCR}); 
this value reads: $\lambda_{\rm crit} \approx a_w d/a^2$.

When we express the parameter $\lambda$ from Eq.~(\ref{eq:32:chiwbwSCRb}) and 
when we substitute it into Eq.~(\ref{eq:32:chiwbwSCRa}), we obtain the 
characteristic function $F(b)$ in the form:
\begin{equation}
    F(b) = 
		\sqrt[3]{b}\, 
		\sqrt[3]{
			\frac{0.67\, P_0^2}{\varepsilon_0^2}\,
			\left(\frac{a_w}{a}\right)^2}.
    \label{eq:33:hSCR}	
\end{equation}
In addition, it is possible to estimate the time evolution of the process of 
charge injection measured by the degree of screening $\lambda(t)$ from the 
time evolution of the parameters $b(t)$ and $\varepsilon_L(t)$:
\begin{equation}
	\label{eq:34:lambdaEvolSCR} 
	\lambda(t) =		 
		\frac{
			\varepsilon_0h \sqrt{\varepsilon_a\varepsilon_c} 
		}{
			1.08\, P_0 d
		}\,
		\sqrt{
			\frac{\varepsilon_L(t)-\varepsilon_c}{b(t)}
		}.
\end{equation}
However, it is seen that the thickness of the passive layer $d$ enters 
into the above equation and its value is always difficult to determine 
experimentally. Instead of that, we can use the expression for the effective 
capacitance per unite area of the ferroelectric-electrode interface 
$C_d\approx \varepsilon_0\sqrt{\varepsilon_a\varepsilon_c}/d$, which can be 
estimated more easily.

\section{Discussion and conclusions}
\label{sec:V}

We have theoretically analyzed the nonlinear dielectric response of the 
ferroelectric polydomain film in two systems with different 
microstructural features: first, the system with the pinned domain walls and, 
second, the system with the electrode-adjacent passive layer. In these two 
considered systems, we studied four possible scenarios of the dielectric aging 
and found that there exists a clear qualitative difference in the evolution of 
the nonlinear dielectric response in each particular scenario. We found 
that the relation between the nonlinearity constant $b_w$ and the small signal 
permittivity $\varepsilon_L$ expressed by the function $F(b_w)
=\chi_w=\varepsilon_L - \varepsilon_c$, which is shown in Tab.~\ref{tab:01:h}, 
is different in the each particular ageing scenario.
\begin{table}[b]
\tabcolsep=4mm 
\caption{
Characteristic functions $F(b)$, which relates the constants of dielectric 
nonlinearity $b$ with the small signal extrinsic contribution to permittivity 
 $\chi_w$ of the polydomain ferroelectric films.
\label{tab:01:h}  }
	\renewcommand{\arraystretch}{2}
\begin{supertabular*}{\textwidth}{ll}
      \hline \hline
         Ageing scenario& $F(b)$ \\
      \hline
         Progressive pinning by crystal lattice defects & 
				$
		\sqrt{b}\, 
		\sqrt{
			\frac{0.031\, P_0^2}{\varepsilon_0^2\varepsilon_c}
			\left(
				\frac{a_w}{a}
			\right)
		}
				$
				\\
         Domain wall coalescence&
				$
		b\,
		\left[
			\frac{0.088\, P_0}{\varepsilon_0\varepsilon_c}
			\left(
				\frac{a_w}{r}
			\right)
		\right]^2
				$ 
				\\
         Thickening of the electrode-adjacent layer& 
				$
		\sqrt[3]{-b}\, 
		\sqrt[3]{
			\frac{0.41\, P_0^2}{\varepsilon_0^2}}
				$
				\\
         Spontaneous polarization screening& 
				$
		\sqrt[3]{b}\, 
		\sqrt[3]{
			\frac{0.67\, P_0^2}{\varepsilon_0^2}\,
			\left(\frac{a_w}{a}\right)^2}
				$
				\\
      \hline \hline
\end{supertabular*}
\end{table}
Therefore, the analysis of the experimentally observed evolution of the 
dielectric nonlinearity constant $b$ and the small signal permittivity 
 $\varepsilon_L$ can be used as a method to identify the microstructural 
mechanism, which plays the dominant role in the evolution of the nonlinear 
dielectric response.

To identify whether the particular microstructural mechanism is dominantly 
responsible for the dielectric ageing, the experimentally measured values of 
 $b$ and $\varepsilon_L$ should be fitted to the function of the general form:
\begin{equation}
    F(b) = K\,\varepsilon_L - B.
    \label{eq:37:hepsfit}	
\end{equation}
If the considered mechanism controls the ageing, the fitted constant $K$ 
should approximately equal to unity and the constant $B$ should be equal to 
 $\varepsilon_c$.

However, applicability of the presented method is subject to several 
considerations that should be kept in mind: (i) We consider that the intrinsic 
permittivity $\varepsilon_c$ is time independent. (ii) We limit our analysis 
to the systems of uniaxial ferroelectric film with the 180${}^\circ$ domain 
pattern and we neglect the effect of the non-180${}^\circ$ domain walls on the 
dielectric response. (iii) We consider that the domain pattern in the 
ferroelectric is neutral, it means that the net spontaneous polarization is 
zero in the absence of the electric field. (iv) We consider that the nonlinear 
dielectric response is dominated by the domain wall movement and not by the 
dielectric response of the crystal lattice, $b_c\ll b_w$. The first assumption 
is intuitive. To justify the second one, we need to point out that the 
extrinsic contribution to the nonlinear permittivity of the non-180${}^\circ$ 
domain walls is negligible due to the elastic effects suppressing their mobility. 
The third assumption puts the restraints on the preparation of the 
ferroelectric samples, which should be depoled, e.g. using a fast decayed low 
frequency ac field. Finally, the last assumption should be checked for 
each ageing scenario and the particular configuration of the domain pattern 
microstructure using the fitted experimental values.

We believe that the presented method can by used as a simple and useful tool 
for getting a deeper insight into the configuration of the domain pattern and 
the quality of the ferroelectric thin films and for providing a way to 
identify the dominant microstructural phenomenon, which is responsible for the 
ageing of dielectric response of ferroelectric polydomain film.

\acknowledgements
This project was supported by the Czech Science Foundation, Project No. 
GACR~202/06/0411. I would like to express my sincere gratitude to Professor 
Jan Fousek for his essential influence on my life, both personal and 
scientific. This work would not be possible without so much important 
experience of sharing his creative ideas, stimulating discussions with 
him, and without receiving his positive feedback. \emph{Thank you!}

\enlargethispage{2em}

\end{document}